\documentclass[a4paper,11pt]{article}

\usepackage{graphicx}
\usepackage{natbib}
\usepackage{here}
\usepackage{setspace}
\usepackage{multirow}
\usepackage{amsmath, amssymb}
\usepackage{bm}
\usepackage{subfigure}
\usepackage{color}

\newcommand{\str}[1]{\renewcommand{\baselinestretch}{#1}\normalsize}

\begin{document}

\str{1.05}

\title
{Nonparametric Bayes models for mixed-scale \\ longitudinal surveys}

\author{\textsc{T. Kunihama} \thanks{Assistant Professor, Department of Economics, Nagoya University, Nagoya, Aichi, Japan. Email: kunihama@soec.nagoya-u.ac.jp} \and \textsc{C. T. Halpern} \thanks{Professor, Department of Maternal and Child Health, Gillings School of Global Public Health, and Carolina Population Center, The University of North Carolina at Chapel Hill, Chapel Hill, North Carolina 27599, U.S.A. Email: carolyn\_halpern@unc.edu} \and \textsc{A. H. Herring} \thanks{Carol Remmer Angle Distinguished Professor of Children's Environmental Health, Department of Biostatistics, Gillings School of Global Public Health, and Carolina Population Center, The University of North Carolina at Chapel Hill, Chapel Hill, North Carolina 27599, U.S.A. Email: aherring@bios.unc.edu}
}

\date{June, 2016}

\maketitle

\vspace{-2mm}


\begin{abstract}

\noindent

Modeling and computation for multivariate longitudinal surveys have proven challenging, particularly when data are not all continuous and Gaussian but contain discrete measurements. In many social science surveys, study participants are selected via complex survey designs such as stratified random sampling, leading to discrepancies between the sample and population, which are further compounded by missing data and loss to follow up.  Survey weights are typically constructed to address these issues, but it is not clear how to include them in models. Motivated by data on sexual development, we propose a novel nonparametric approach for mixed-scale longitudinal data in surveys.  In the proposed approach, the mixed-scale multivariate response is expressed through an underlying continuous variable with dynamic latent factors inducing time-varying associations.  Bias from the survey design is adjusted for in posterior computation relying on a Markov chain Monte Carlo algorithm.  The approach is assessed in simulation studies, and applied to the National Longitudinal Study of Adolescent to Adult Health.
\\


\end{abstract}


\str{1.55}

\section{Introduction}

The National Longitudinal Study of Adolescent to Adult Health (Add Health) has collected vast amounts of information about social, economic and health-related behaviours in adolescence over 20 years.  
One of the primary aims is to study trajectories of adolescent sexual development into adulthood.  In particular, it is of interest to investigate associations among different indicators of sexual orientation, including adolescent sexual behaviours, attraction and identity; changes in the associations over the transition to adulthood, and whether those changes may vary by demographic characteristics. Although Add Health data include all the sexual development indicators of interest, several characteristics of the data make it challenging to build a realistic statistical model.

First, the data consist of mixed-scale variables, e.g. partner counts or nominal sexual orientation identity.  It is not straightforward to model mixed-scale multivariate data jointly, especially if they include both ordered variables and nominal variables.  Second, individuals repeatedly answered a set of questions over time, leading to subject-specific time-dependence.   Third, the Add Health study drew its sample via a complex survey design with stratified sampling of schools and oversampling of numerous groups.  Hence, the data are not representative of the population until sampling weights are applied.  Fourth, it is not efficient to build a statistical model separately for each subpopulation, because the number of respondents in certain groups can be relatively small, especially when considering weight adjustments that account for loss to follow-up.  In order to make stable population-representative inferences and use small area estimation, it is essential to include information for all subpopulations.  Fifth, missing values are common, and we observe both design-based and individual-specific missingness.  For example, students were not asked as detailed questions on their sexuality in adolescence as in young adulthood by design.  Therefore, we need to conduct statistical analysis of adolescent sexual development, taking fully into account all of these challenging characteristics of the Add Health data.

There is a rich literature on modeling of mixed-scale data.  One approach is to apply generalized linear models for each outcome in which dependence among variables is induced through shared latent factors (\cite{SammelRyanLegler97}; \cite{Dunson00}; \cite{MoustakiKnott00}; \cite{DunsonHerring05}).  However, the robustness of the approaches based on generalised linear mixed models can be weak due to the dual role of the random effects structure in controlling the dependence and shape of marginal distributions.  Another approach is to use underlying continuous variables, specified by a Gaussian model (\cite{Muthen84}) or a Dirichlet process mixture model (\cite{KottasMullerQuintana05}; \cite{CanaleDunson11}; \cite{KimRatchford13}).  In this approach, discrete variables can be expressed by thresholding the latent continuous variables.  Avoiding specification of marginal distributions, \cite{Hoff07} proposed a semiparametric Gaussian copula model in which associations among mixed-scale variables are induced by correlations among the latent Gaussian variables.  \cite{MurrayDunsonCarinLucas13} and \cite{GruhlEroshevaCrane13} extend this approach by incorporating factor structures but these copula models can incorporate only ordered variables.  Also, \cite{MurrayReiter14} propose a nonparametric Bayesian joint model for multiple imputation of missing values.  \cite{McParland-et-al14} develop a model-based clustering approach for mixed-scale data that combines item response theory models for ordered variables with factor analysis models for nominal variables.  

Recently, a number of articles have addressed the analysis of multivariate longitudinal data (\cite{Bandyopadhyay-et-al11}, \cite{Verbeke-et-al14}).  \cite{Dunson03} proposed dynamic latent trait models in which autoregressive Gaussian latent factors incorporate subject-specific time-dependence and generalized linear models describe mixed-scale outcomes.  It is routine to incorporate time effects into multivariate models through time-varying covariates, such as polynomial functions of age, with random coefficients (\cite{GueorguievaSanacora06}; \cite{FieuwsVerbeke06}; \cite{LuoWang14}; \cite{Baghfalakia-et-al14}).  \cite{Liu-et-al09} develop a joint model for longitudinal binary and continuous variables, consisting of a correlated probit model and a regression model with the Bartlett decomposition of a covariance matrix.  \cite{GhoshHanson10} propose a semiparametric approach with a mixture of Polya trees for random effect distributions.  \cite{DasDaniels14} develop a semiparametric model for bivariate sparse longitudinal data, applying a matrix stick-breaking process for a residual covariance matrix.  However, it is not clear how to handle survey bias, missingness and small area estimation in these approaches.  Therefore, none of these methods can capture the exact nature of longitudinal surveys in Add Health.

In the literature on survey data analysis, two major approaches are used: the design-based approach, which treats outcomes as fixed quantities and the model-based approach, which models outcomes and effectively predicts values for the non-sampled subjects in a population (\cite{Little04}; \cite{LevyLemeshow08}; \cite{Rao11}).  \cite{Little04} and \cite{Gelman07} show difficulties with current methods in practice and highlight the importance of including survey weights in model-based analyses.  \cite{ZhengLittle03, ZhengLittle05} propose a nonparametric method that flexibly models the outcome given inclusion probability using a penalized spline, and  \cite{ChenElliottLittle10} extend the model for binary variables.  \cite{SiPillaiGelman15} propose a nonparametric Bayesian model, which jointly models an outcome and survey weights based on a Gaussian process regression.  However, these approaches are developed mainly for inference of the population mean of a univariate response, and extensions to mixed-scale longitudinal data are not straightforward.

We propose a flexible nonparametric model for the analysis of Add Health sexual orientation development data.  In the proposed approach, mixed-scale variables are expressed through transformation of latent continuous variables, for which a Dirichlet process mixture of Gaussian factor models is developed.  For unordered categorical variables, we employ the concept of utilities in multinomial probit models, in which the nominal outcome is a manifestation of underlying continuous utility variables.  The subject-specific dynamic variability can be captured by time-varying latent factors via Gaussian processes which can easily incorporate irregular time intervals for each respondent.  Inferences valid for the source population are obtained by adjusting the mixture weights in the Dirichlet process mixtures using the survey weights.  Because we build a joint model of the response variables and covariates, missing values can be easily imputed assuming missing at randomness.  Also, from the proposed joint density, we construct a conditional density on covariates for small area inferences.  For posterior computation, we develop an efficient Markov chain Monte Carlo (MCMC) algorithm in which we modify the Dirichlet process mixture with survey weights, taking into account uncertainty in the adjustment process.

Section \ref{sec:data} describes our data set from Add Health.  Section \ref{sec:modeling} proposes a novel approach for mixed-scale longitudinal surveys.  Section \ref{sec:mcmc} develops an efficient MCMC algorithm.  Section \ref{sec:simulation} assesses the performance of the proposed approach against competitors through simulation.  Section \ref{sec:analysis} applies the proposed method to the adolescent sexual development questions of interest.  Section 7 concludes the article.

\section{National Longitudinal Study of Adolescent to Adult Health (Add Health) data} \label{sec:data}

Add Health is a nation-wide longitudinal study of adolescents in grades 7 to 12 in the United States in the 1994-95 academic year (http://www.cpc.unc.edu/projects/addhealth).  The cohort study has been following up respondents with four waves of in-home interviews conducted in 1994-95, 1996, 2001-02, and 2008-09.  It collects data on respondents' social, behavioral, psychological and biological information, allowing researchers to study developmental trajectories across the life course of adolescence into adulthood.

The Add Health study design selects a stratified sample of 80 high schools and 52 middle schools from the United States with unequal probability of selection based on region, urbanicity, school type, ethnic mix, and size.  Then, students in each school are stratified by grade and sex.  In addition, the study drew supplemental samples, oversampling groups of particular interest based on ethnicity, genetic relatedness to siblings, adoption status, disability and parental education.  Taking into account non-response as well as the sample selection, survey weights have been constructed for population-representative inferences via weighting adjustments.  The full study design is described by \cite{Harris09}.  

\begin{figure}[htpb] 
\begin{center}
\includegraphics[width=12cm,height=6cm]{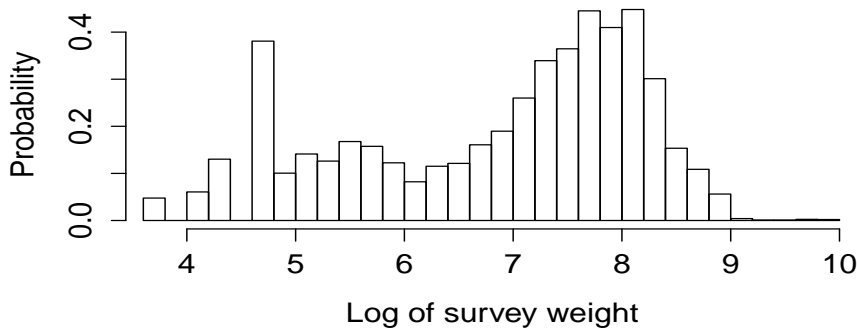}
\caption{Histogram of the logarithm of the survey weight.}
\label{fig:log-weight}
\end{center}
\end{figure}

To understand adolescent sexual orientation development, our primary goal is studying associations among measures of sexual attraction, behavior, and identity in adolescence and trends in the transition to adulthood.  We use three waves of surveys which were conducted when participants were in grades 7-12 in wave 1, young adults in wave 3 and adults in wave 4.  Table \ref{tb:list} shows the list of variables in our data set.  They are mixed-scale, and the sexual orientation identity is missing by design in wave 1.  For longitudinal studies with these three waves, survey weights are constructed for $n=12,288$ respondents. Figure \ref{fig:log-weight} reports the histogram of the weights on a logarithmic scale.  

\begin{figure}[htpb] 
\begin{center}
\vspace{-5mm}
\includegraphics[width=16cm,height=10.3cm]{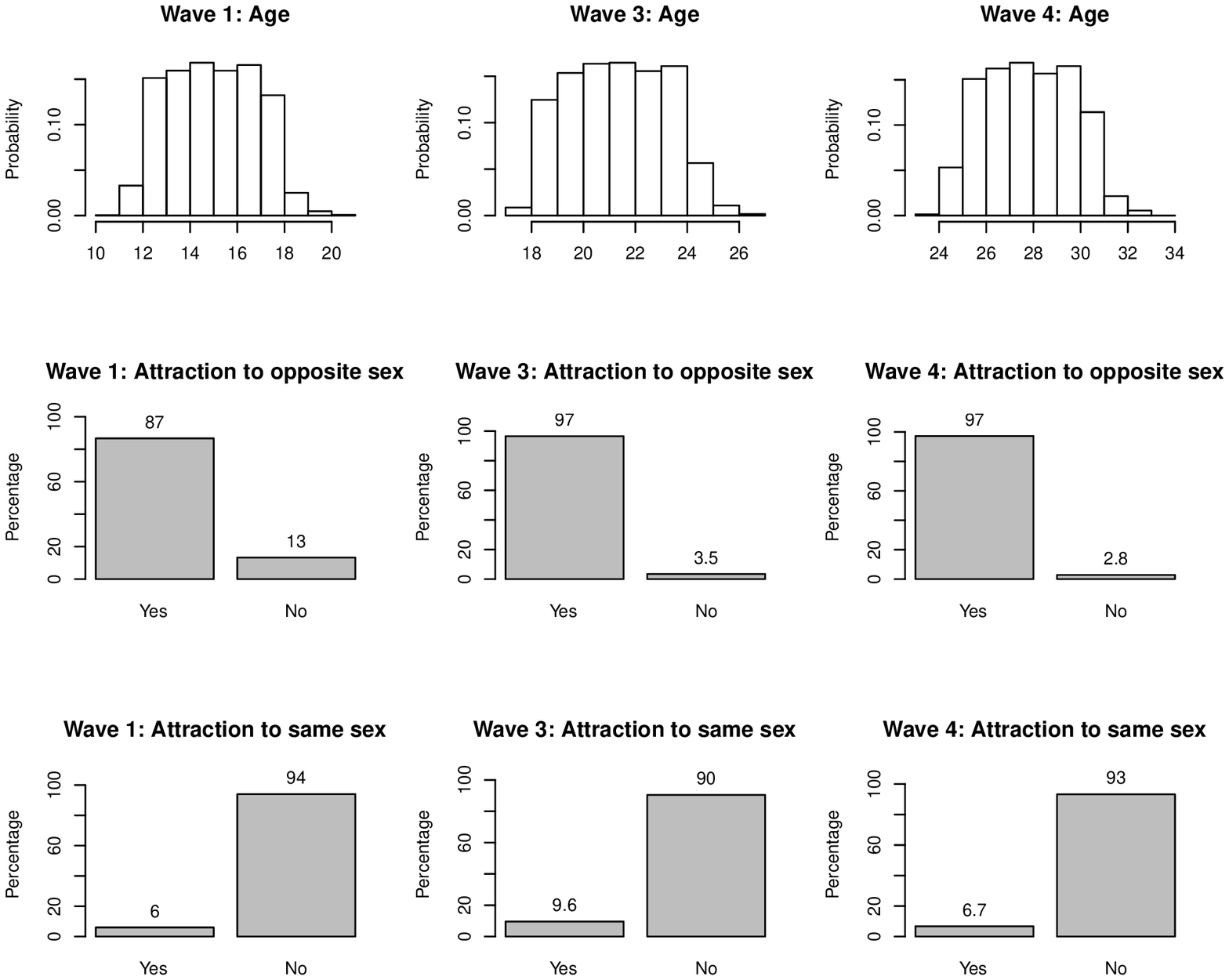}
\includegraphics[width=16cm,height=10.3cm]{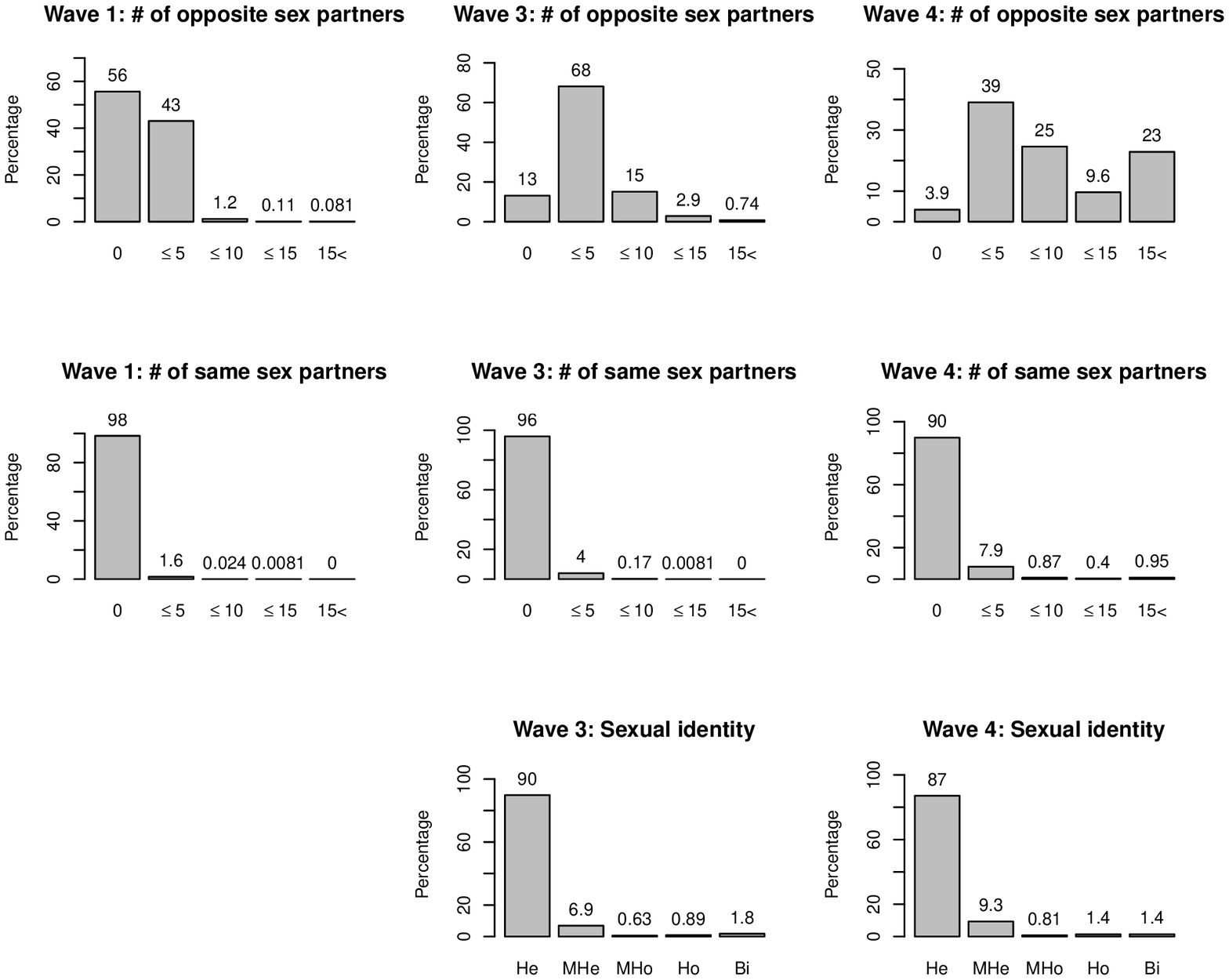} 
\vspace{-5mm}
\caption{Histogram of age and response variables in wave 1 (left), 3 (middle) and 4 (right). In sex partners, $\leq 5$, $\leq 10$, $\leq 15$ mean 1 to 5, 6 to 10, 11 to 15. In sexual identity, He, MHe, MHo, Ho, Bi represent heterosexual, mostly heterosexual, mostly homosexual, homosexual and bisexual.}
\label{fig:data-plot}
\end{center}
\end{figure}

\begin{figure}[htpb] 
\begin{center}
\includegraphics[width=15.5cm,height=5cm]{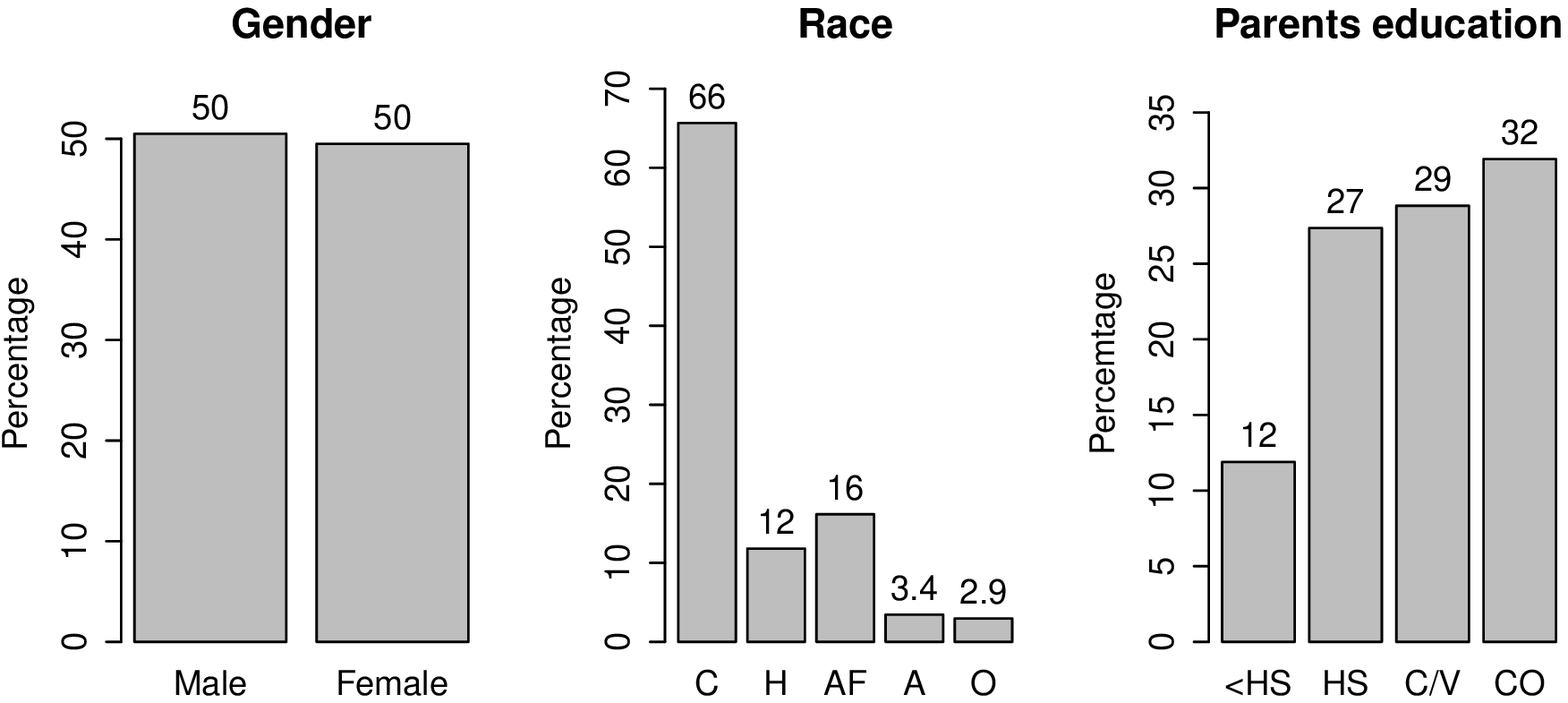}
\includegraphics[width=15.5cm,height=5cm]{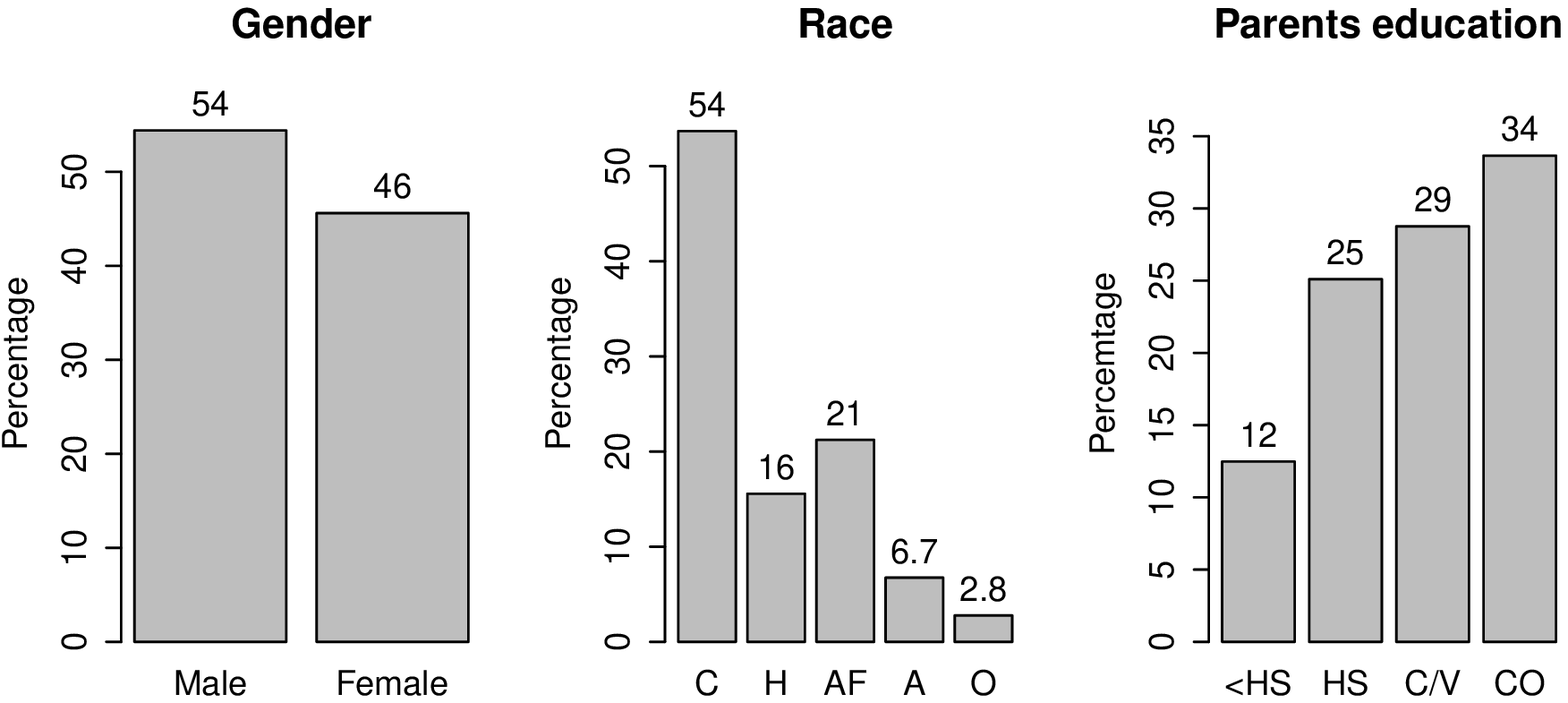}
\caption{Histogram of the weighted (above) and unweighted (below) covariates. In race, C, H, AF, A, O represent Caucasian, Hispanic, African American, Asian, Other race. In parents education, $<$HS, HS, C/V, Co correspond to less than high school, high school, college/voc, college graduate.}
\label{fig:covariate-plot}
\end{center}
\end{figure}

Figure \ref{fig:data-plot} plots the weighted sexual development variables and age in each wave.  Although the total numbers of sex partners are categorized into five groups for the display, we treat them as count data in the real data analysis in Chapter 6.  To obtain the weighted data, we simply resampled $n=12,288$ respondents from the original data set with probability $\tilde{w}_i=w_i / \sum_{j=1}^n w_j$ where $w_i$ is the survey weight for $i$th respondent.  Figure \ref{fig:covariate-plot} reports the plot of the weighted and unweighted covariates, indicating the gap induced by the sampling design. For example, racial minorities are oversampled in the survey, leading to the relatively large decreased percentage of Caucasian in the raw data. Although the resampling method works for computing basic summary statistics, it may be inefficient in terms of discarding information of the unselected respondents in the original data set. Also, it is not clear how to incorporate uncertainty about the unsampled subjects in the population. We discuss an alternative adjustment method in Section \ref{sec:survey-weight}, which takes fully into account the information of all respondents in our data set and reflects uncertainty about the unobserved subjects in the population.
 
\begin{table}[t]
\centering
\small
\begin{tabular}{lll}
\hline
	& Type  &  Categories \\
\hline 
{\bf Sexual development variables} & & \\
Attraction to opposite sex  	&	binary	 & 0. No, 1. Yes \\
Attraction to same sex  		&	binary	 & 0. No, 1. Yes \\
Total \# of opposite sex partners   	&	count	 &  \\
Total \# of same sex partners   		&	count	 &  \\
Sexual identity    	&	nominal	 & 1. Heterosexual, 2. Mostly Heterosexual, \\
(missing at wave 1) & & 3. Mostly Homosexual, 4. Homosexual, 5. Bisexual  \\
& & \\
{\bf Covariates} & & \\
Gender  &	binary	 & 0. Male, 1. Female \\
Race  	&	nominal	 & 1. Caucasian, 2. Hispanic, 3. African American, \\
& & 4. Asian, 5. Other race \\
Parental education   &	ordered	 & 1. Less than high school, 2. High school diploma/GED,  \\
& categorical & 3. Some college/voc education, 4. College graduate+ \\
\hline
\end{tabular}
\normalsize
\caption{List of sexual development variables and covariates.}
\label{tb:list}
\end{table}

\section{Proposed modeling of mixed-scale longitudinal surveys} \label{sec:modeling}

\subsection{Modeling of mixed-scale data with latent continuous variables} \label{sec:function}

Let $y=(y_1,\ldots,y_p)'$ be a multivariate response with mixed-scale margins.  We introduce a latent continuous variable $y^*=(y_1^*,\ldots,y^*_{p})'\in \mathbb{R}^{p}$ and express the response $y$ by transforming the underlying variable $y^*$.  For example, a binary variable $y_k \in \{0,1\}$ can be induced by 
\begin{align*}
\text{binary}: \hspace{0.5cm} y_k = 1 (y^*_{k} > 0),
\end{align*}
where $1(\cdot)$ denotes an indicator function.  A count variable $y_k \in \{0,\cdots,\infty\}$ can be expressed as
\begin{align*}
\text{count}: \hspace{0.5cm} y_k = \sum_{r=0}^{\infty} r \times 1 ( a_{r} < y_{k}^* \leq a_{r+1}),
\end{align*}
where $-\infty = a_0 < a_1 < a_2 < \cdots<\infty$.  For a nominal variable with $d$ categories, we translate it to $d$ indicator variables $y_{k_1},\ldots,y_{k_d}$ with $y_{k_l}\in\{0,1\}$ for $l=1,\ldots,d$ and $\sum_{l=1}^d y_{k_l}=1$.  Then, we set
\begin{align}
\text{nominal}: \hspace{0.5cm} y_{k_l} =  1( y^*_{k_l} = \max_{1 \leq l' \leq d} y^*_{k_{l'}} ), \ \ l=1,\ldots,d.
\label{eq:unordered}
\end{align}
In the equation (\ref{eq:unordered}), $y^*_{k_l}$ can be interpreted as utility for the $l$th category as in multinomial probit models (\cite{McCullochRossi94}; \cite{ImaiDyk05}; \cite{BurgetteNordheim12}).  Because the order of $y_{k_1}^*,\ldots,y^*_{k_d}$ is unchanged with respect to adding any constant and multiplying any positive scale by all of them, without loss of generality we may assume one of the utilities is zero.

\subsection{Proposed nonparametric Bayes model} \label{sec:proposed}

We apply the above framework to mixed-scale longitudinal data.  Let $y_{ij}=(y_{ij1},\ldots,y_{ijp})'$ be the response variable for respondent $i$ at age $t_{ij}$ for $i=1,\ldots,n$ and $j=1,\ldots,n_i$ with $y_{ijk}\in \mathcal{Y}_k$ allowed to be any type of univariate random variable for $k=1,\ldots,p$.  We introduce $y^*_{ij}=(y^*_{ij1},\ldots,y^*_{ijp})'\in \mathbb{R}^{p}$ as a latent continuous variable, which induces $y_{ij}$ through a function $g$ such that $y_{ij} = g(y^*_{ij})$ as in Section \ref{sec:function}.  Also, $x_{i}=(x_{i1},\ldots,x_{iL})' \in \mathcal{X}$ denotes static covariates for $i$th respondent, from which target subpopulations may be constructed.  

We propose a novel nonparametric Bayes model, in which time-varying factors flexibly capture a dynamic structure of the response by
incorporating time effects and covariates into its mean and covariance.  Let $\eta_{ij}=(\eta_{ij1},\ldots,\eta_{ijQ})'$ denote dynamic factors for respondent $i$ at age $t_{ij}$ with $Q < p$.  We first consider modeling of the latent response variable given the covariates,
\begin{align}
y^*_{ij} = B x_{i} + \Omega \mu_{t_{ij}} + \Lambda \eta_{ij} + \epsilon_{ij}, \ \epsilon_{ij} \sim N(0, \Sigma),  \ \
\Sigma =\text{diag}(\sigma^2_{1},\ldots,\sigma^2_{p}),
\label{eq:y-x} 
\end{align}
where $\mu_t=(\mu_{t1},\ldots,\mu_{tQ_{\mu}})'$ is a dynamic effect for age $t$ with $Q_{\mu}<p$, $B$ is a $p\times L$ matrix of regression coefficients, $\Omega$ is a $p \times Q_{\mu}$ coefficient matrix for the time effect, and $\Lambda$ is a $p \times Q$ factor loading matrix.  In this model (\ref{eq:y-x}), we can capture age-specific effects for all individuals through $\mu_t$ and additional subject-specific time variability is induced by the dynamic factors $\eta_{ij}$.  Relying on the covariance regression by \cite{HoffNiu12} and \cite{FoxDunson15}, we assume that the dynamic factors can be expressed as
\begin{align}
\eta_{ij} = Vx_i \eta^*_i + \xi_{t_{ij}} \tilde{\eta}_i, \ \ \eta^*_i\sim N(0, 1), \ \ \tilde{\eta}_i\sim N(0, I_{Q_{\eta}}),
\label{eq:eta}
\end{align}
where $V$ is a $Q\times L$ coefficient matrix, $\xi_t=\{\xi_{tql}\}$ is a $Q\times Q_{\eta}$ matrix for $t=1,\ldots,T$ and $\eta^*_i$ and $\tilde{\eta}_i$ are random effects.  Integrating out the random effects, we obtain a covariance varying with covariates and age,
\begin{align}
y^*_{ij} \sim N(B x_{i} + \Omega \mu_{t_{ij}}, \ \Lambda Vx_i x'_i V' \Lambda' + \Lambda \xi_{t_{ij}} \xi'_{t_{ij}} \Lambda' + \Sigma).
\label{eq:y}
\end{align}
Let $\mu_{\bullet q}=\{ \mu_{tq}, \ t\in \mathcal{T} \}$ and $\xi_{\bullet ql}=\{ \xi_{t ql}, \ t\in \mathcal{T} \}$ be stochastic processes of the time effects with time space $\mathcal{T}$, for which we apply Gaussian processes,
\begin{align}
\mu_{\bullet q} &\sim \text{GP}(0,c_{\mu}), \ c_{\mu}(\mu_{tq},\mu_{t'q}) = \text{Cov}(\mu_{tq},\mu_{t'q}) = \exp (-\kappa_{\mu} | t - t' |^2 ), \\
\xi_{\bullet ql} &\sim \text{GP}(0,c_{\xi}), \ c_{\xi}(\xi_{tql},\xi_{t'ql}) = \text{Cov}(\xi_{tql},\xi_{t'ql}) = \exp (-\kappa_{\xi} | t - t' |^2 ),
\end{align}
where $\kappa_{\mu}>0$, $\kappa_{\xi}>0$ and $\mu_{\bullet q}$ and $\xi_{\bullet ql}$ are mutually independent with respect to $q$ and $l$.  

Next, relying on the conditional density of the latent variable (\ref{eq:y-x})-(\ref{eq:y}), we develop a flexible joint model of the response and covariates.  The approach of the response-covariate joint modeling has been widely studied in the literature (\cite{MullerErkanliWest96}; \cite{MullerQuintana10}; \cite{HannahBleiPowell11}; \cite{TaddyKottas12}; \cite{DeYoreoKottas15b, DeYoreoKottas15a}).  In this framework, we can derive a marginal density $f(y)$ and a conditional density $f(y\,|\,x\in C)$ given $C \subset \mathcal{X}$ with covariate space $\mathcal{X}$ from the joint model $f(y,x)$.  Also, we can easily impute missing values from the conditional distribution given observed information assuming missing at randomness.  

We construct a joint density of $y^*_i$ and $x_i$ with $y^*_{i}=\{y^*_{ij}, \ j=1,\ldots,n_i\}$, relying on Dirichlet process mixtures (\cite{Lo84}; \cite{WestMullerEscobar94}; \cite{EscobarWest95}),
\begin{align}
f(y^*_{i}, x_i) &= \int f(y^*_{i}\,|\, x_{i}, \theta^y) f(x_{i}\,|\, \theta^x) dP(\theta), \ \ P \sim \text{DP}(\alpha P_0),
\label{eq:joint-0} \\ 
&= \sum_{h=1}^{\infty} \pi_h f(y^*_{i}\,|\, x_{i}, \theta^y_h ) \prod_{l=1}^L f_l(x_{il}\,|\,\theta^x_h),
\label{eq:joint-1} \\ 
\pi_h &= v_h \prod_{b<h} (1 - v_b), \ \ v_h \sim \text{Beta}(1,\alpha),
\label{eq:joint-2}
\end{align}
where $P$ denotes a prior distribution for $\theta = (\theta^y, \theta^x)$, $\alpha>0$ is a precision parameter and $P_0$ is a base probability measure for the Dirichlet process, and $f(y^*_{i}\,|\, x_{i}, \theta^y_h)$ denotes the conditional density induced by (\ref{eq:y-x})-(\ref{eq:y}).   With respect to the density for the covariate $f_l(x_{il}\,|\,\theta^x_h)$ in (\ref{eq:joint-1}), we assume a function depending on the type of the covariate, such as a multinomial distribution for a categorical variable.  Equations (\ref{eq:joint-1}) and (\ref{eq:joint-2}) correspond to the stick-breaking representation of the Dirichlet process mixtures (\cite{Sethuraman94}; \cite{MuliereTardella98}).  The proposed model is the Dirichlet process mixture of Gaussian factor models.  Finally, integrating out the latent variable over the space derived from the function in Section \ref{sec:function}, we obtain the joint density of the response and covariates,
\begin{align}
f(y_{i}, x_i) &= \int_{R(y_{i})} f(y^*_{i}, x_i ) d y^*_{i}, 
\end{align}
where $R(y_{i})=\{ y_{i}^* \in \mathbb{R}^{p \times n_i}:  \ y_{ij} = g(y_{ij}^*), \ j=1,\ldots,n_i \}$. 

\subsection{Population-representative inference on scale-free association} \label{sec:survey-weight}

To adjust for the sample bias from complex survey designs, we incorporate survey weights into the proposed joint model.  Let $D$ be the finite population with $N$ people, from which $n$ subjects are sampled with inclusion probability $\nu_i$ for subject $i\in D$.  Inverse of the inclusion probability $\nu_i^{-1}$ can be interpreted as the number of people the respondent $i$ represents in the population (\cite{HorvitzThompson52}). Also, the survey weights can be defined as $w_i \propto \nu_i^{-1}$.  For the finite approximation of the Dirichlet process mixtures with $H$ components, \cite{Kunihama-etal16} develop an adjustment method, which modifies the mixture weights $\{\pi_h\}$ in (\ref{eq:joint-1}) using the survey weights. In the posterior computation, along with parameter update, we generate adjusted mixture weights $\tilde{\pi}=(\tilde{\pi}_1,\ldots,\tilde{\pi}_H)'$ from
\begin{align}
\tilde{\pi} \sim \text{Dirichlet}\left( a_1 +  \sum_{i:s_i=1} \frac{w_i}{c}, \ldots, a_H + \sum_{i:s_i=H} \frac{w_i}{c} \right),
\label{eq:step}
\end{align}
where $c = \sum_{i=1}^n w_i / N$ and $s_i$ is the cluster index variable for the $i$th respondent.  With the prior distribution $\text{Dirichlet}(a_1\ldots,a_H)$, the posterior distribution (\ref{eq:step}) consists of the prior sample size, which represents uncertainty about the unsampled subjects in the population, and the enlarged observations based on the survey weights.  As a default setting, the prior sample size is 1-2\% of population size $N$.  Compared to the posterior distribution $\text{Dirichlet}(a_1+\sum_{i}1(s_i=1),\ldots,a_H+\sum_i 1(s_i=H))$ in a standard Bayesian mixture model, the weight of subject $i$ is changed from 1 to $w_i/c \approx\nu_i^{-1}$.

Let $\tilde{f}(y,x)$ be the adjusted joint density with the mixture weights $\tilde{\pi}$ from (\ref{eq:step}).  Relying on the adjusted density $\tilde{f}$, we estimate associations of mixed-scale variables by generating samples from the posterior predictive distributions.  As a scale-free measure of dependence between two variables, we employ Goodman and Kruskal's gamma (\cite{GoodmanKruskall54};  \cite{GoodmanKruskall59}; \cite{GoodmanKruskall63}; \cite{GoodmanKruskall72}),
\begin{align}
\gamma = \frac{N_c - N_d}{N_c + N_d},
\label{eq:gk-gamma}
\end{align}
where $N_c$ is the number of concordant pairs, and $N_d$ is the number of discordant pairs.  A concordant pair can be defined as a pair of variables ($X_1$,$Y_1$) and ($X_2$,$Y_2$) such that $\text{sgn}(X_2-X_1)=\text{sgn}(Y_2-Y_1)$.  On the other hand, a discordant pair means $\text{sgn}(X_2-X_1)=-\text{sgn}(Y_2-Y_1)$. The measure $\gamma$ ranges from -1 (100\% negative association) to 1 (100\% positive association), and zero indicates the absence of association. Although other popular rank correlations such as Spearman's rho need an adjustment for the ties, which are neither concordant nor discordant pairs, Goodman and Kruskal's gamma is resistant to them.  

To estimate associations for the subpopulation defined by $C \subset \mathcal{X}$, we generate $\tilde{y}_r$ from $\tilde{f}(y\,|\,x \in C) = \tilde{f}(y, x\in C)/\tilde{f}(x\in C)$ with $r=1,\ldots,R$ where $\tilde{y}_r=\{\tilde{y}_{rt}, \ t\in \mathcal{T} \}$ and $\tilde{y}_{rt}=(\tilde{y}_{rt1},\ldots,\tilde{y}_{rtp})'$ is a mixed-scale response for time $t$. Then, we compute the concordant and discordant pairs for $\tilde{y}_{rtj}$ and $\tilde{y}_{rtj'}$ with $j\neq j'$ and calculate the Goodman and Kruskal's gamma through (\ref{eq:gk-gamma}) for each time point.  For the overall population, we can estimate the gamma similarly by setting $C=\mathcal{X}$.

\section{Posterior computation} \label{sec:mcmc}

Relying on the blocked Gibbs sampler by \cite{IshwaranJames01}, we develop an efficient MCMC algorithm for the proposed model.  Letting $U_{h} = [B_{h} \ \Omega_{h} \ \Lambda_{h}]$ for the $h$th mixture component, we apply shrinkage priors for it with high density around zero to reduce effects of redundant elements while the tails are heavy enough to capture important signals.  We assume $U_{hkl}\sim N(0, \delta^2_{kl})$, $\delta^2_{kl} \sim \text{IG}(0.5,0.5)$ where IG denotes an inverse-gamma distribution for $k=1,\ldots,p$ and $l=1,\ldots, L^*$ with $L^*=L+Q_{\mu}+Q$.  By integrating out $\delta^2_{kl}$, the prior corresponds to a Cauchy prior which is a commonly-used shrinkage prior with heavy tails.  We also apply the same type of shrinkage prior to $V_h$, that is, $V_{hql}\sim N(0, \zeta^2_{ql})$, $\zeta^2_{ql} \sim \text{IG}(0.5,0.5)$ for $q=1,\ldots,Q$ and $l=1,\ldots, L$.  As a notation below, let $A_{k\bullet}$, $A_{\bullet l}$ and $a_{(-k)}$ be the $k$th row and $l$th column of a matrix $A$ and a subvector of a vector $a$ excluding $k$th component.  Also, $\hat{y}_{ij}$ denotes a standardized latent response, which is differently defined at a different step.  Then, we propose the following MCMC algorithm. 
  
\begin{enumerate}
\item Update $v_h$ from Beta($1 + \sum_{i=1}^n 1(s_i=h)$, $\alpha+ \sum_{i=1}^n 1(s_i > h)$) for $h=1,\ldots,H-1$.

\item Using the prior distribution $\text{Gamma}(a_{\alpha}, b_{\alpha})$, update $\alpha$ from $\text{Gamma}(a_{\alpha}+H-1, b_{\alpha} - \sum_{h=1}^{H-1} \log(1-v_h))$.

\item Update $s_{i}$ for $i=1,\ldots,n$ from  
\begin{align*}
P(s_{i}=h\,|\, \cdots) &= \frac{ \pi_h \prod_{j=1}^{n_i} f(y^*_{ij}\,|\, x_{i}, \eta_{ij}, \theta^y_h ) \prod_{l=1}^L f_l(x_{il} \,|\, \theta^x_{h}) }{ \sum_{m=1}^{H} \pi_m \prod_{j=1}^{n_i} f(y^*_{ij}\,|\, x_{i}, \eta_{ij}, \theta^y_m ) \prod_{l=1}^L f_l(x_{il} \,|\, \theta^x_{m}) }. 
\end{align*}

\item Using the prior $\text{IG}(\tilde{a}_{k},\tilde{b}_{k})$, update $\sigma^2_{hk}$ for $k=1,\ldots,p$ and $h=1,\ldots,H$ from 
\begin{align*}
\text{IG} \left( \frac{\sum_{i:s_i=h}n_i + \tilde{a}_{k}}{2}, \frac{\sum_{i:s_{i}=h} \sum_{j=1}^{n_i} \hat{y}_{ijk}^2 + \tilde{b}_{k}}{2} \right),
\end{align*}
where $\hat{y}_{ijk} = y_{ijk}^*-B_{hk\bullet} x_i - \Omega_{hk\bullet} \mu_{t_{ij}} - \Lambda_{hk\bullet} \eta_{ij}$.

\item Update $U_{hk\bullet}$ for $k=1,\ldots,p$ and $h=1,\ldots,H$ from $N(\mu_{U},\Sigma_{U})$ where
\begin{align*}
\mu_{U} = \Sigma_{U} \left( \sigma^{-2}_{hk} \sum_{i:s_i=h} \sum_{j=1}^{n_i} z_{ij} y^*_{ijk} \right), \ \ \Sigma_{U} = \left( \sigma^{-2}_{hk} \sum_{i:s_i=h} \sum_{j=1}^{n_i} z_{ij} z'_{ij} + \Sigma^{-1}_{0} \right)^{-1},
\end{align*}
where $z_{ij}= (x_i', \mu'_{t_{ij}}, \eta'_{ij})'$ and $\Sigma_{0} = \text{diag}(\delta^2_{k1},\ldots,\delta^2_{kL^*})$.

\item Update $\eta^*_{i}$ from $N(\mu_{\eta^*}, \sigma^2_{\eta^*})$ for $i=1,\ldots,n$ where
\begin{align*}
\mu_{\eta^*} = \sigma^2_{\eta^*} x'_i V'_{s_i} \Lambda'_{s_i} \Sigma_{s_i}^{-1} \sum_{j=1}^{n_i} \hat{y}_{ij}, \ \ 
\sigma^2_{\eta^*} = \left( n_i x'_i V'_{s_i} \Lambda'_{s_i} \Sigma_{s_i}^{-1} \Lambda_{s_i} V_{s_i} x_i + 1 \right)^{-1},
\end{align*}
where $\hat{y}_{ij}=y^*_{ij}-B_{s_i} x_i - \Omega_{s_i} \mu_{t_{ij}} - \Lambda_{s_i} \xi_{t_{ij}} \tilde{\eta}_i$.

\item Update $\tilde{\eta}_i$ from $N(\mu_{\tilde{\eta}}, \Sigma_{\tilde{\eta}})$ for $i=1,\ldots,n$ where
\begin{align*}
\mu_{\tilde{\eta}} = \Sigma_{\tilde{\eta}} \left( \sum_{j=1}^{n_i}  \xi'_{t_{ij}} \Lambda'_{s_i} \Sigma_{s_i}^{-1} \hat{y}_{ij} \right), \ \ \Sigma_{\tilde{\eta}} = \left( \sum_{j=1}^{n_i} \xi'_{t_{ij}} \Lambda'_{s_i} \Sigma_{s_i}^{-1} \Lambda_{s_i} \xi_{t_{ij}} + I \right)^{-1},
\end{align*}
with $\hat{y}_{ij}=y^*_{ij}-B_{s_i} x_i - \Omega_{s_i} \mu_{t_{ij}} - \Lambda_{s_i} V_{s_i} x_i \eta^*_i$.

\item Using the Griddy-Gibbs sampler by \cite{RitterTanner92}, update $\kappa_{\mu}$ from
\begin{align*}
P(\kappa_{\mu} = c_k \,|\,\cdots)= \frac{\prod_{q=1}^{Q_{\mu}} f_N(\mu_{\bullet q} \,|\, 0, \Psi(c_{k}))}{\sum_{l=1}^G \prod_{q=1}^{Q_{\mu}} f_N(\mu_{\bullet q} \,|\, 0, \Psi(c_{l}))},  
\end{align*}
where $c_1,\ldots,c_G$ are grid points and $f_N(\cdot\,|\,0,\Psi(c_{k}))$ denotes a normal density with mean 0 and covariance $\Psi(c_{k})$, which is derived from the Gaussian process with the length-scale $c_k$.
 
\item Using the Griddy-Gibbs sampler, update $\kappa_{\xi}$ from
\begin{align*}
P(\kappa_{\xi} = c_k \,|\,\cdots)= \frac{\prod_{q=1}^{Q}\prod_{l=1}^{Q_{\eta}} f_N(\xi_{\bullet ql} \,|\, 0, \Psi(c_{k}))}{\sum_{j=1}^G \prod_{q=1}^{Q} \prod_{l=1}^{Q_{\eta}} f_N(\xi_{\bullet ql} \,|\, 0, \Psi(c_{j}))}.  
\end{align*}
 
\item Update $V_{h\bullet l}$ for $h=1,\ldots,H$ and $l=1,\ldots,L$ from $N(\mu_{V},\Sigma_{V})$ with
\begin{align*}
\mu_{V} = \Sigma_{V}  \Lambda'_{h} \Sigma_{h}^{-1} \left( \sum_{i:s_i=h} x_{il} \eta^*_i \sum_{j=1}^{n_i} \hat{y}_{ij}  \right), \ \ 
\Sigma_{V} = \left( \sum_{i:s_i=h} n_i x^2_{il} \eta^{*2}_i \Lambda'_{h} \Sigma_h^{-1} \Lambda_{h} + \Sigma_{0V}^{-1} \right)^{-1},
\end{align*}
where $\hat{y}_{ij} = y^*_{ij} - B_{h} x_{i} -\Omega_{h} \mu_{t_{ij}} - \Lambda_{h} \xi_{t_{ij}} \tilde{\eta}_i - \Lambda_{h} V_{h\bullet (-l)} x_{i(-l)} \eta^*_i $ and $\Sigma_{0V}=\text{diag}(\zeta^2_{1l},\ldots\zeta^2_{Ql})$

\item Update $\mu_{\bullet q}=(\mu_{1q},\ldots,\mu_{Tq})'$ for $q=1,\ldots,Q_{\mu}$ from $N(\mu_{\mu},\Sigma_{\mu})$ with $\mu_{\mu} = \Sigma_{\mu} b$, $\Sigma^{-1}_{\mu} = A + \Psi^{-1}(\kappa_{\mu})$, $A= \text{diag}(a_{1},\ldots,a_{T})$, $b=( b_{1},\ldots,b_{T} )'$ and 
\begin{align*}
a_{t} = \sum_{(i,j):t_{ij}=t} \Omega'_{s_i\bullet q} \Sigma_{s_i}^{-1} \Omega_{s_i\bullet q}, \ \ 
b_{t}=\sum_{(i,j):t_{ij}=t} \Omega'_{s_i\bullet q} \Sigma^{-1}_{s_i} \hat{y}_{ij}, 
\end{align*}
where $\hat{y}_{ij} = y^*_{ij} - B_{s_i} x_{i} - \Lambda_{s_i} \eta_{ij} - \Omega_{s_i\bullet (-q)} \mu_{t_{ij}(-q)}$.

\item Update $\xi_{\bullet ql}=(\xi_{1ql},\ldots,\xi_{Tql})'$ for $q=1,\ldots,Q$ and $l=1,\ldots,Q_{\eta}$ from $N(\mu_{\xi},\Sigma_{\xi})$ with $\mu_{\xi} = \Sigma_{\xi} b^*$ and $\Sigma^{-1}_{\xi} = A^* + \Psi^{-1}(\kappa_{\xi})$, $A^*= \text{diag}(a^*_{1},\ldots,a^*_{T})$, $b^*=( b^*_{1},\ldots,b^*_{T} )'$ and 
\begin{align*}
a^*_{t} = \sum_{(i,j):t_{ij}=t} \Lambda_{s_i\bullet q} \Sigma^{-1}_{s_i} \Lambda_{s_i\bullet q} \tilde{\eta}^{2}_{il}, \ \ 
b^*_{t}= \sum_{(i,j):t_{ij}=t}  \Lambda'_{s_i\bullet q} \Sigma^{-1}_{s_i} \hat{y}_{ij} \tilde{\eta}_{il}, 
\end{align*}
where $\hat{y}_{ij} = y^*_{ij} - B_{s_i} x_{i} - \Omega_{s_i} \mu_{t_{ij}} - \Lambda_{s_i} V_{s_i} x_i \eta^*_i - c^*$ with $c^*=(c^*_{1},\ldots,c^*_{p})'$ and $c^*_{k} = \sum_{(q^*,l^*)\neq (q,l)} \Lambda'_{s_i k q^*} \xi_{tq^*l^*} \tilde{\eta}_{il^*}$.

\item Update $\theta^x_{hl}$ in $\theta^x_h=(\theta^x_{h1},\ldots,\theta^x_{hL})'$ for $h=1,\ldots,H$ and $l=1,\ldots,L$ from
\begin{align*}
f(\theta^x_{hl}\,|\,\cdots) \propto \prod_{i:s_i=h} f_l(x_{il} \,|\, \theta^x_{hl}) \pi(\theta^x_{hl}),
\end{align*}
where $\pi(\theta^x_{hl})$ is the prior density.

\item Impute missing values $x_{il}$ from
\begin{align*}
f(x_{il} \,|\, \cdots) \propto \prod_{j=1}^{n_i} f(y^*_{ij}\,|\, x_{i}, \eta_{ij}, \theta^y_{s_i} ) f_l(x_{il} \,|\, \theta^x_{s_i l}).
\end{align*}

\item Impute missing values $y_{ijk}$ by generating $y_{ijk}=g(y_{ijk}^*)$, $y^*_{ijk} \sim f(y^*_{ijk}\,|\,x_i, \eta_{ij}, \theta^y_{s_i})$. 

\item Update $y^*_{ijk}$ for $i=1,\ldots,n$, $j=1,\ldots,n_i$ and $k=1,\ldots,p$ from 
\begin{align*}
f(y^*_{ijk}\,|\,\cdots) \propto 1\{y_{ij}=g(y^*_{ij})\} f(y^*_{ijk}\,|\,x_i, \eta_{ij}, \theta^y_{s_i}).
\end{align*}

\item Update $\delta^2_{kl}$ from IG($0.5(H+1)$, $0.5(\sum_{h=1}^H U^2_{hkl}+1)$) for $k=1,\ldots,p$ and $l=1,\ldots,L^*$.

\item Update $\zeta^2_{ql}$ from IG($0.5(H+1)$, $0.5(\sum_{h=1}^H V^2_{hql}+1)$) for $q=1,\ldots,Q$ and $l=1,\ldots,L$. 

\item Generate the adjusted mixture weights $\tilde{\pi}$ and compute the Goodman and Kruskal's gamma as in Section \ref{sec:survey-weight}.
\end{enumerate}

\section{Simulation study} \label{sec:simulation}

We assess performance of the proposed method with simulation data by comparing to two existing approaches.  We assume that the response consists of continuous, binary, count and three-category nominal variables, and the covariate is a binary variable.  All respondents participate in the survey three times out of $T=9$ time points, randomly taking $t_{ij} \in \{3j-2, 3j-1, 3j\}$ with equal probabilities for $j=1,2,3$.  We estimate associations in the mixed-scale response for groups with $x_i=0$ and $x_i=1$.  For the priors in the proposed method, we use $\alpha\sim \text{Gamma}(0.25,0.25)$, $\sigma^2_{hk}\sim \text{IG}(2,\tilde{s}_k/200)$ where $\tilde{s}_k$ is the sample variance of the $k$th response for the ordered variables and 1 for the nominal variable.  Also, we set $H=60$ and $Q=Q_{\mu}=Q_{\eta}=4$, and put 25 grids on the $(0,1]$ interval for $\kappa_{\mu}$ and $\kappa_{\xi}$.  

For competitors, we consider two methods. The first one is based on covariance regression with a polynomial function of time (\cite{HoffNiu12}), 
\begin{align*}
y_{ij}=\tilde{g}(y^*_{ij}), \ \ y^*_{ij}=Az_{ij}+Bz_{ij} \eta_i +\varepsilon_{ij}, \ \ \eta_i\sim N(0,1), \ \varepsilon_{ij}\sim N(0,\Psi), 
\end{align*}
where $z_{ij} = (1,x_i,t_{ij},t_{ij}^2,t_{ij}^3)'$ and $A$ and $B$ are $p\times 5$ coefficient matrices.  The function $\tilde{g}$ transforms the latent variable $y^*_{ij}$ with the identity map for a continuous variable, the probit function for a binary variable, discretization with unknown cut-points for a count variable and the multinomial probit function for a nominal variable.  For the prior distributions, we assume $\Psi\sim \text{Inverse-Wishart}(\Psi^{-1}_0,\nu_0)$ with expectation $\Psi_0/(\nu_0-p-1)$ and $(A, B)\,|\,\Psi\sim \text{Matrix-Normal}(0,\Psi,V_0)$ where $\nu_0=p+2$, $\Psi_0$ is the sample covariance matrix of $y_{ij}$ and $V_0$ is a block diagonal matrix where the two blocks are $n (Z'Z)^{-1}$ with $Z=(z_1,\ldots,z_n)'$ and $z_i=(z_{i1}, z_{i2}, z_{i3})$.   Also, we apply independent prior $N(0, 100)$ to thresholds for a count variable.  

The second approach is a dynamic latent trait model (\cite{Dunson03}), 
\begin{align*}
y_{ij} \sim f(y_{ij}\,|\,\zeta_{ij}), \  \zeta_{ij} = A \tilde{x}_{i} + \lambda_{t_{ij}} \tau_{it_{ij}} + \phi_{i}, \ \tau_{i \bullet}\sim\text{GP}(\beta \tilde{x}_i, c_{\tau}), \ \phi_{i}\sim N(0,\Phi),
\end{align*}
where $\tilde{x}_i=(1,x_i)'$, $A$ is a $p\times 2$ matrix, $\lambda_t$ is a factor loading vector at time $t$, $\tau_{i\bullet}=\{\tau_{it}, t\in \mathcal{T}\}$, $c_{\tau}(\tau_{it},\tau_{it'})=\exp(\kappa_{\tau}|t-t'|^2)$ and $\Phi=\text{diag}(\omega_1^2,\ldots,\omega_p^2)$.  For the density function $f$, we assume Gaussian regression with variance $\sigma^2$ for continuous, logistic regression for binary, Poisson regression for count and multinomial logistic regression for nominal variables.  As for the priors, we assume $\sigma^2\sim \text{IG}(2, \tilde{s}/2)$, $\tilde{s}$ is the sample variance of the continuous response, $A \sim \text{Matrix-Normal}(0, I, 100 I)$, $\lambda_{t} \sim N(0, 100I)$, $\omega_k^2\sim \text{IG}(2, 5)$, $\beta \sim N(0, 100I)$ and discrete grid prior with 25 points on $(0, 1]$ for $\kappa_{\tau}$.  For all methods, we generated 10,000 samples after the initial 5,000 samples are discarded as a burn-in period and every 10th sample is saved.  To compute the Goodman and Kruskal's gamma, we generated $4,000$ sample from the posterior predictive distribution.  We observed that the sample paths were stable and the sample autocorrelations dropped smoothly.  

\begin{figure}[htbp]
 \begin{minipage}{0.5\hsize}
  \begin{center}
   \includegraphics[width=80mm]{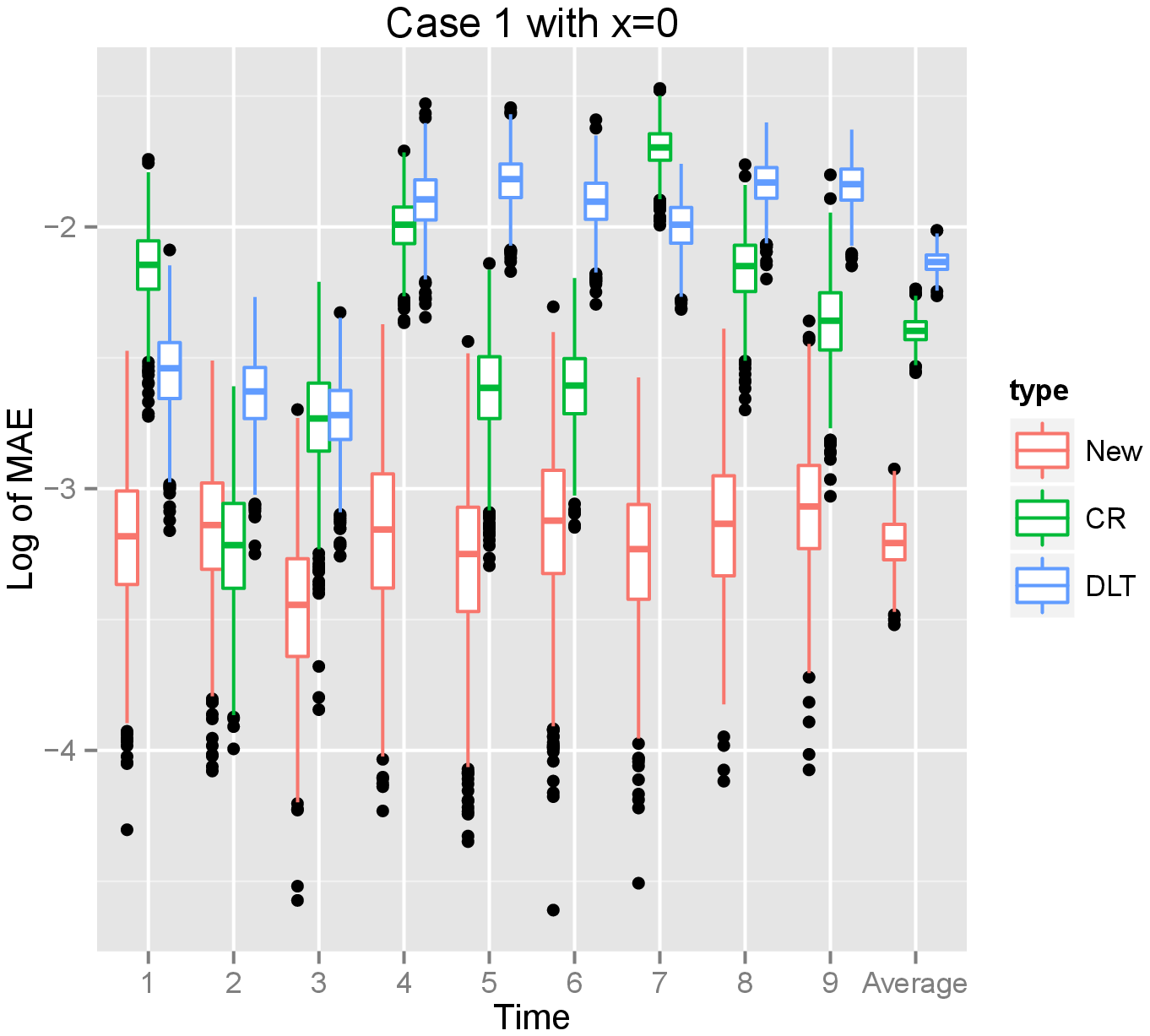}
  \end{center}
  \label{fig:one}
 \end{minipage}
 \begin{minipage}{0.5\hsize}
  \begin{center}
   \includegraphics[width=80mm]{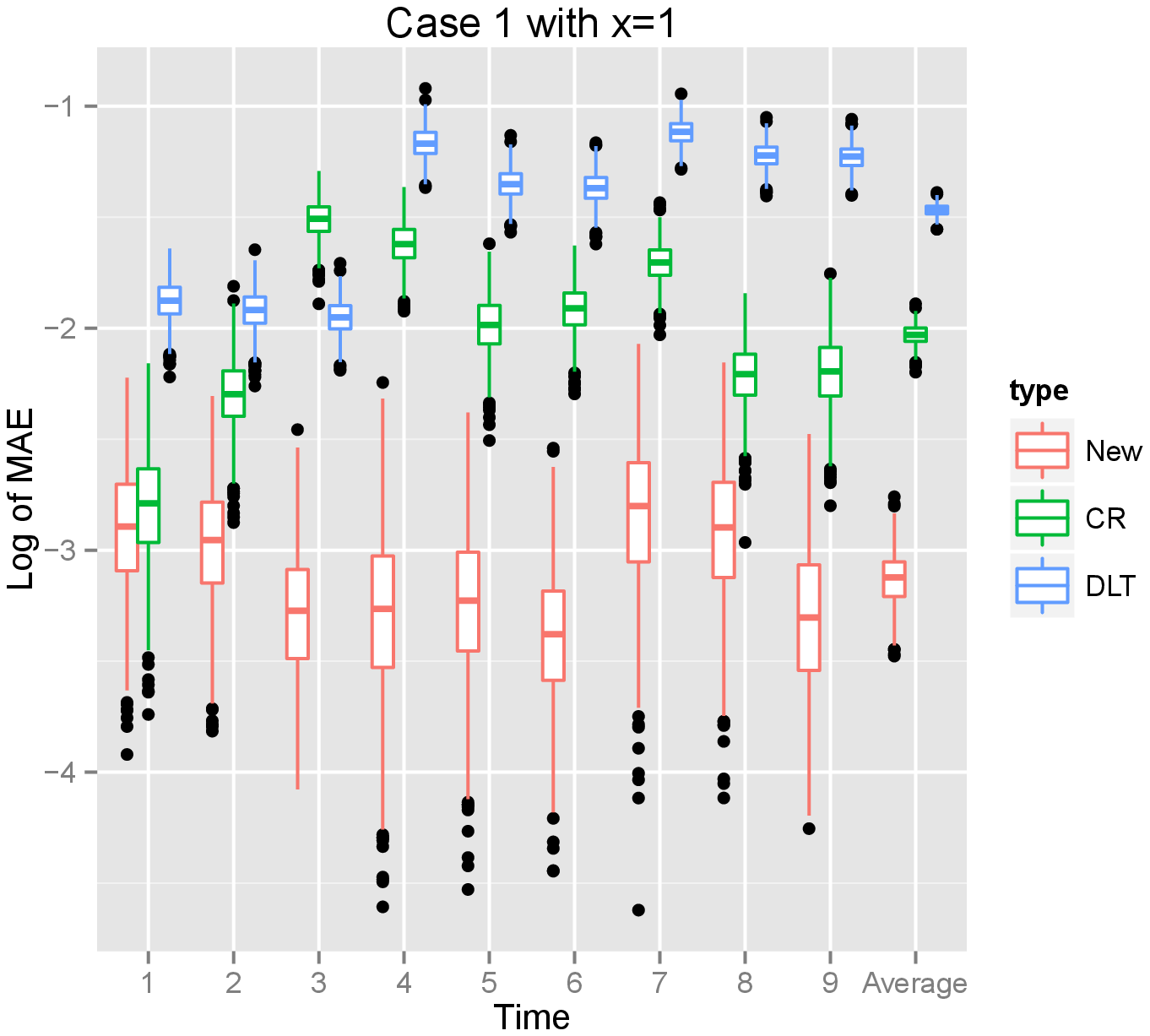}
  \end{center}
  \label{fig:two}
 \end{minipage}
\caption{Boxplots of logarithm of mean absolute errors for the proposed model (New), covariance regression (CR) and dynamic latent trait model (DLT) in case 1 with $x=0$ (left) and $x=1$ (right). The $x$-axis shows time points and the average.}
\label{fig:case-1}
\end{figure}

In case 1, we generated data with $n=4,000$ subjects from
\begin{align*}
y_{ij}&=g(y^*_{ij}), \ y^*_{ij}=D \tilde{x}_i + F_{t_{ij}} \tilde{x}_i \eta_i + \varepsilon_{ij}, \ \varepsilon_{ij}\sim N(0, 0.1^2I), \\
\eta_i &\sim N(0,1), \ \tilde{x}_i=(1,x_i)', \ x_i\sim \text{Bernoulli}(0.5),
\end{align*}
where each element of $D$ is independently generated from $N(0, 0.05^2)$. We assume $F_{t}=F_{t-1}+\varepsilon^*_t$, $\varepsilon^*_t\sim \text{Matrix-Normal}(0, I, 0.05^2 I)$ for $t=4,7$ and $F_t=F_{t-1}$ for the other time points with $F_0\sim \text{Matrix-Normal}(0, I, 0.05^2 I)$.  These matrices are given in the supplementary materials.  The function $g$ corresponds to the one in Section \ref{sec:function} with the identity map for the continuous variable and the zero utility of the last category for the nominal variable.  Also, we use non-negative integers as cutting points for the count variable.  Because it is not straightforward to analytically compute true values of the Goodman and Kruskal's gamma, we approximate them by generating 8,000 samples from the true data-generating function.  Figure \ref{fig:case-1} reports boxplots of logarithm of mean absolute errors at each time point and averaged ones over time in case 1.  Although the covariance regression model shows close values at some time points, the proposed method constantly produces the smallest errors over time in both groups $x=0$ and $x=1$. Also, the proposed method reports the lowest averaged errors over the period of time with relatively large differences.  

Next, we simulated data with $n=4,000$ subjects from the model in which time-dependence is induced by the covariance regression with a second-order polynomial function of time, and the sample is divided into two subgroups.
\begin{align*}
y_{ij}&=g(y^*_{ij}), \ y^*_{ij}=D_{s_i} z_{ij} + F_{s_{i}} z_{ij} \eta_i + \varepsilon_{ij}, \ \varepsilon_{ij}\sim N(0, \sigma_{s_i}^2 I), \ \eta_i\sim N(0,1), \\
z_{ij}&=(1,x_i, \tilde{t}_{ij}, \tilde{t}_{ij}^2)', \ x_i\sim \text{Bernoulli}(0.5), \  s_i\sim \text{Bernoulli}(0.5),  
\end{align*}
where $\tilde{t}=t/T$ and $\sigma_0=0.1$, $\sigma_1=0.05$. Each component of $D_s$ and $F_s$ is independently generated from $N(0, 0.05^2)$ with $s=0,1$ and given in the supplementary materials. Figure \ref{fig:case-2} shows the estimation result of the logarithm of the mean absolute errors of associations for $x=0$ and $x=1$ in case 2.  The covariance regression model reports smaller errors over time than the dynamic latent trait model, but the proposed method outperforms the competitors in both $x=0$ and $x=1$.

\begin{figure}[htbp]
 \begin{minipage}{0.5\hsize}
  \begin{center}
   \includegraphics[width=80mm]{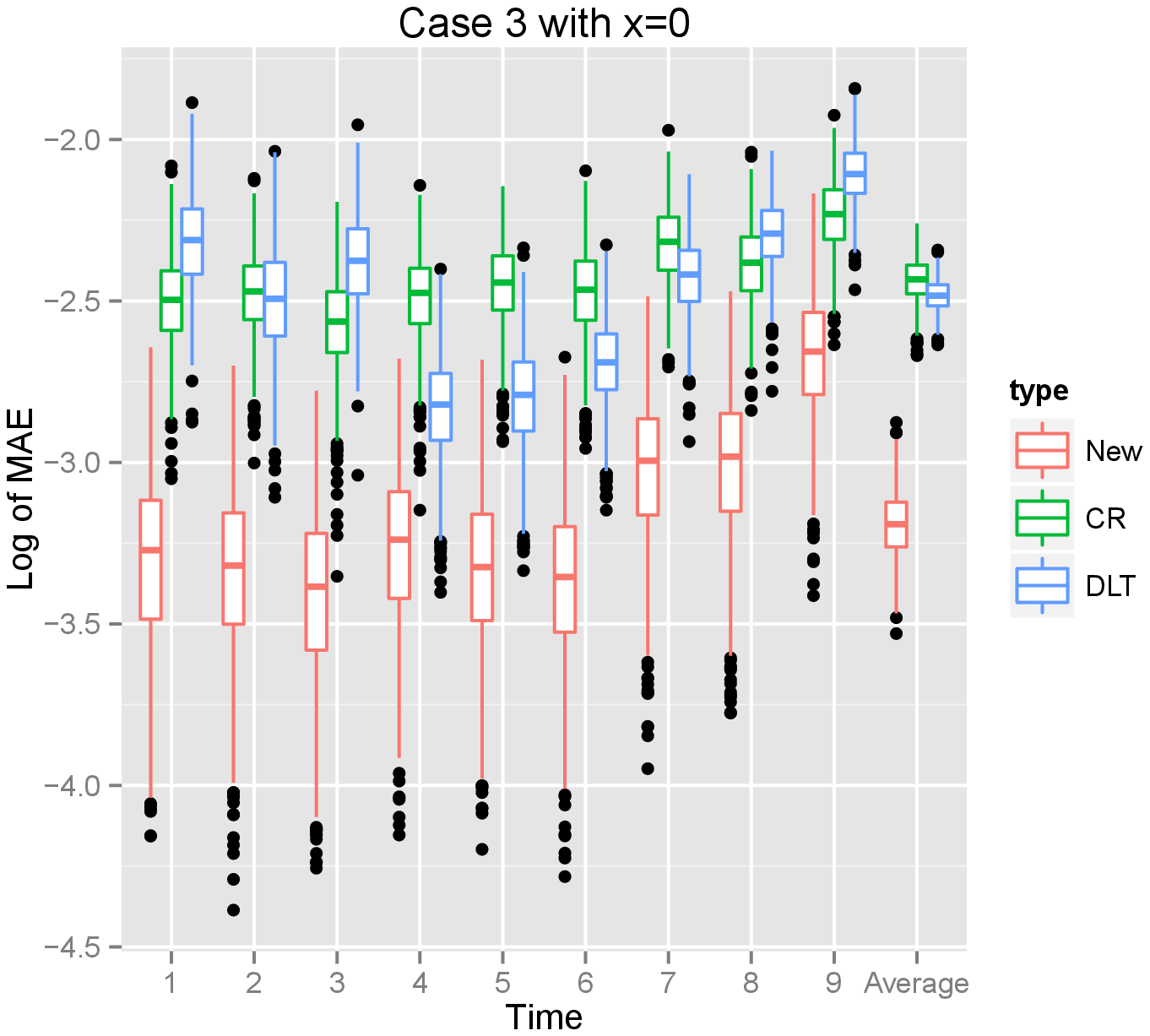}
  \end{center}
  \label{fig:one}
 \end{minipage}
 \begin{minipage}{0.5\hsize}
  \begin{center}
   \includegraphics[width=80mm]{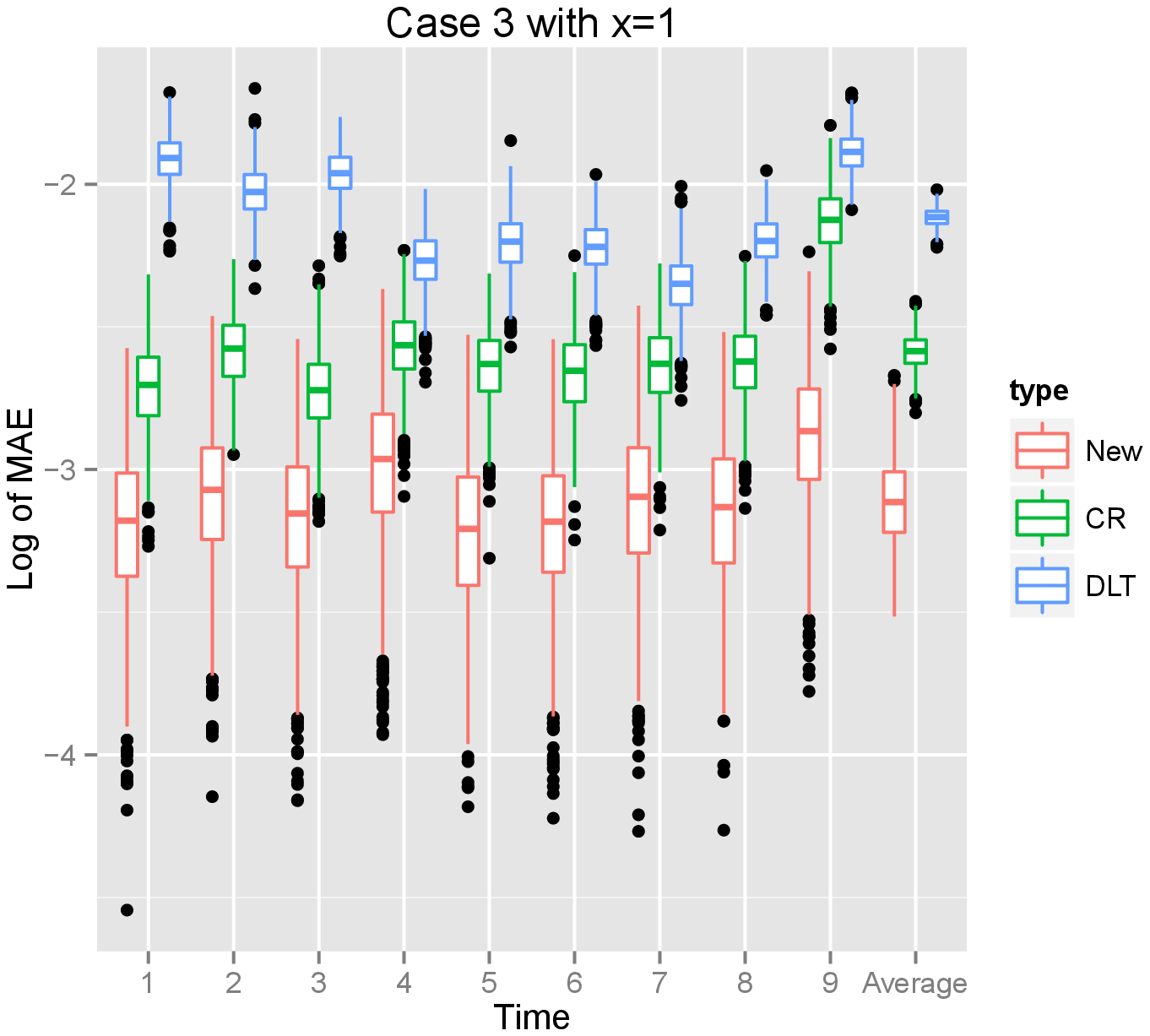}
  \end{center}
  \label{fig:two}
 \end{minipage}
\caption{Boxplots of logarithm of mean absolute errors for the proposed model (New), covariance regression (CR) and dynamic latent trait model (DLT) in case 3 with $x=0$ (left) and $x=1$ (right). The $x$-axis shows time points and the average.}
\label{fig:case-3}
\end{figure}

Finally, we assume data are collected in a stratified sampling design.  We consider a finite population with $N=1,000,000$ subjects consists of three subpopulations having $N_1=650,000$, $N_2=300,000$ and $N_3=50,000$ people.  We sample $n_m=1,500$ subjects from each stratum with $m=1,2,3$ and construct survey weights by $w_i=N_{m_i}/n_{m_i}$ for $i$th subject where $m_i$ represents the subpopulation to which $i$th subject belongs.  Then, we assume the following data-generating function, expressing dynamic structures through the covariance regression as in case 2.
\begin{align*}
y_{ij}&=g(y^*_{ij}), \ y^*_{ij}=D_{m_i} z_{ij} + F_{m_{i}} z_{ij} \eta_i + \varepsilon_{ij}, \ \varepsilon_{ij}\sim N(0, 0.1^2 I), \ \eta_i\sim N(0,1), \\
z_{ij}&=(1,x_i, \tilde{t}_{ij}, \tilde{t}_{ij}^2)', \ x_i\sim \text{Bernoulli}(0.5), 
\end{align*}
where all component in $D_m$  and $F_m$ are independently generated from $N(0, 0.05^2)$ with $m=1,2,3$ and given in the supplementary materials. In the proposed method, we set the prior sample size as 1\% of population size $N$ for the bias adjustment in Section \ref{sec:survey-weight}. The estimation result of the log mean absolute errors is given in Figure \ref{fig:case-3} in case 3. In both cases with $x=0$ and $x=1$, the proposed method reports the smallest mean absolute errors at each time point and the lowest averaged value over time.

\begin{figure}[htbp]
 \begin{minipage}{0.5\hsize}
  \begin{center}
   \includegraphics[width=80mm]{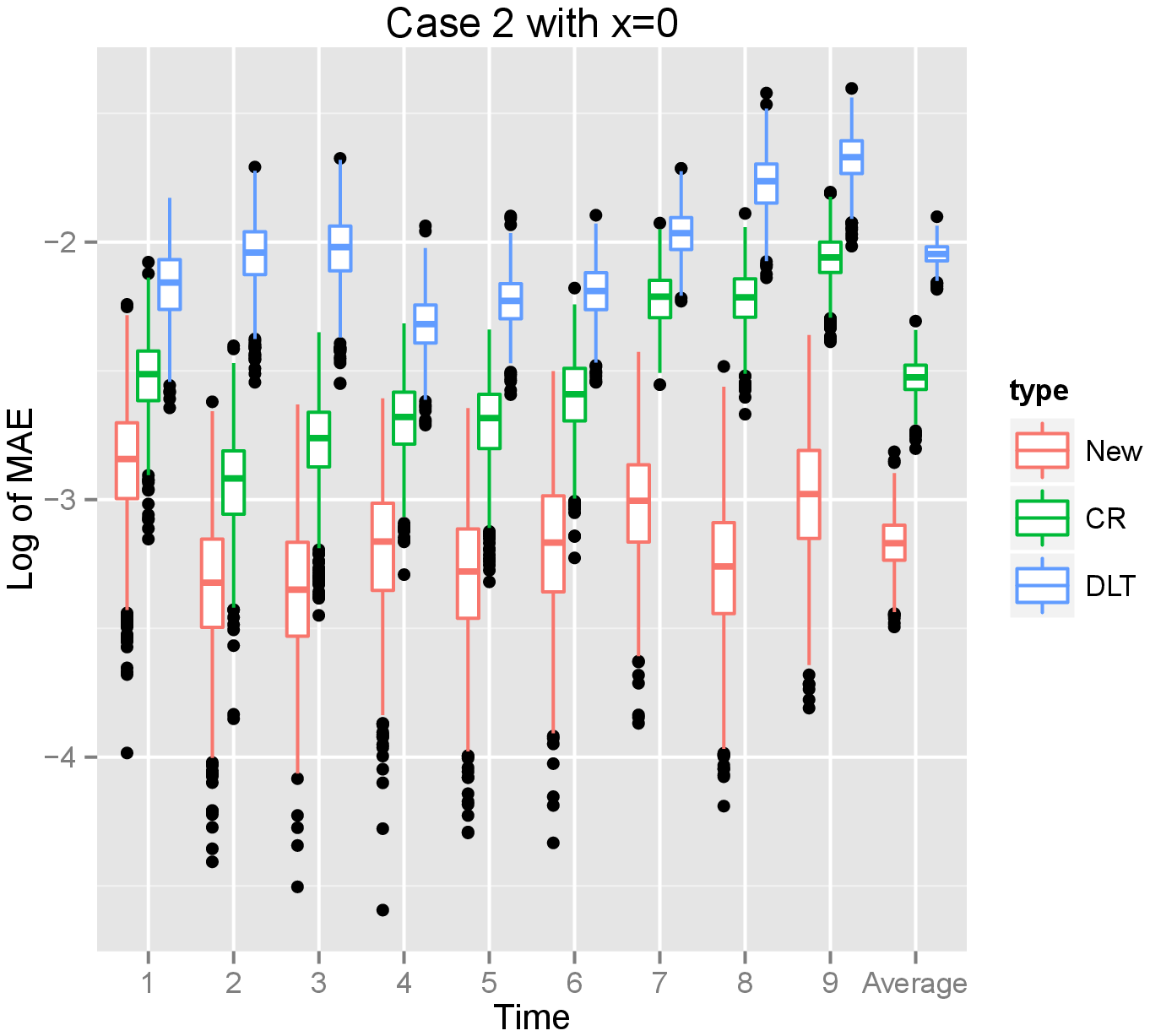}
  \end{center}
  \label{fig:one}
 \end{minipage}
 \begin{minipage}{0.5\hsize}
  \begin{center}
   \includegraphics[width=80mm]{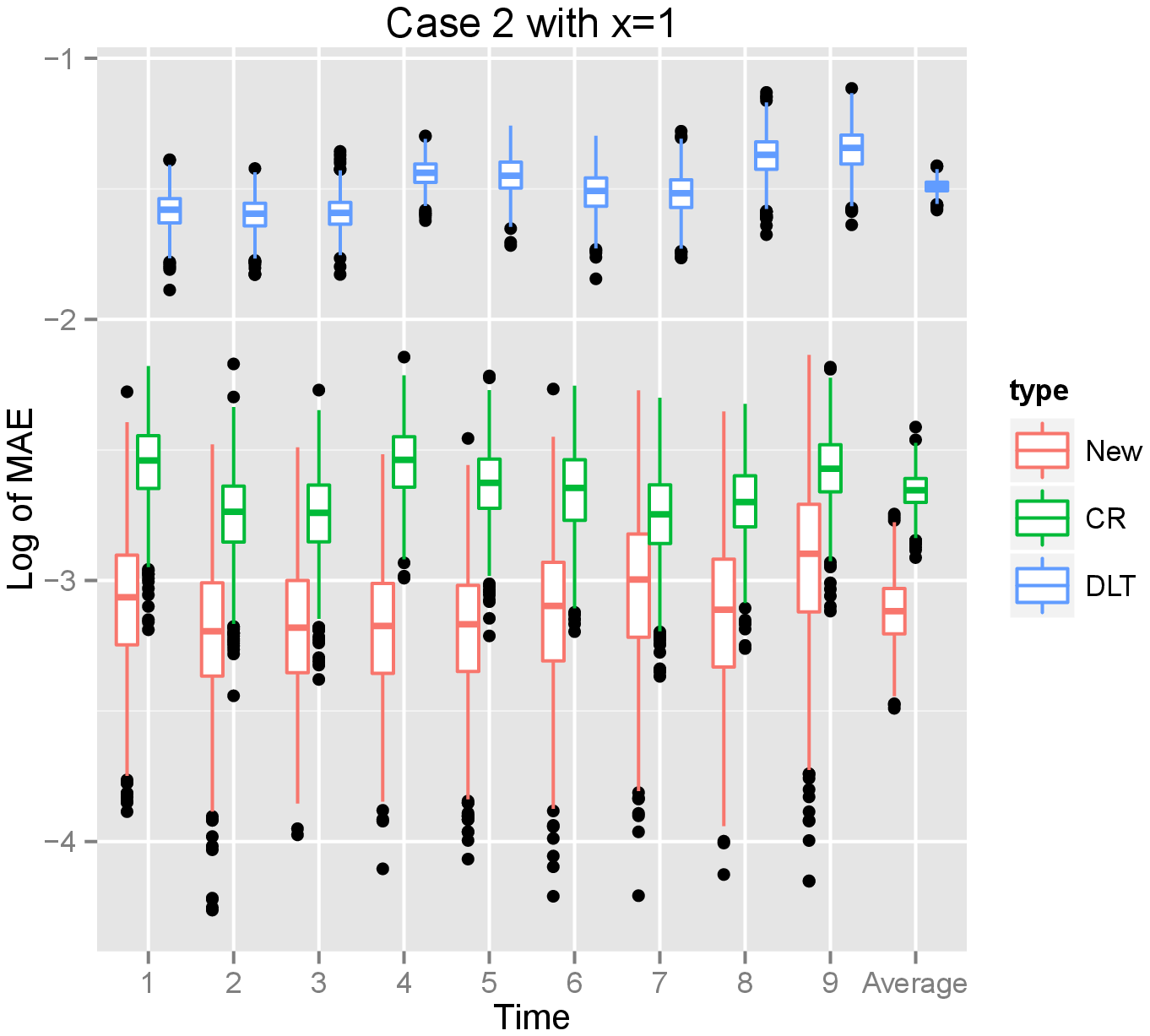}
  \end{center}
  \label{fig:two}
 \end{minipage}
\caption{Boxplots of logarithm of mean absolute errors for the proposed model (New), covariance regression (CR) and dynamic latent trait model (DLT) in case 2 with $x=0$ (left) and $x=1$ (right). The $x$-axis shows time points and the average.}
\label{fig:case-2}
\end{figure}

\section{Analysis of adolecent sexual development data} \label{sec:analysis}

This section applies the proposed method to the adolescent sexual development data. As for the function $f_l(x_{il}\,|\,\theta^x)$ for the covariates in (\ref{eq:joint-1}), we assume multinomial distributions using Dirichlet priors with all concentration parameters 1.  For the count variables, because we observe high right skewness in these data, we use log cut-points instead of nonnegative integers, which allow the Dirichlet process mixtures to efficiently approximate such distributions.  Also, as the prior of the count variables, we assume $\sigma^2_{hk}\sim \text{Inverse-Gamma}(2,\tilde{s}_k/200)$ where $\tilde{s}_k$ is the sample variance of $\log(y_{ijk}+0.5)$.  For the other parameters, we use the same prior settings as in the simulation study in Section \ref{sec:simulation}.  In the sampling bias adjustment, we set the hyperparameter in the Dirichlet prior such that the prior sample size is equal to 1\% of the population size $N=14,677,347$. We estimated trajectories of associations among sexual development variables for the overall population and within subpopulations with respect to gender, race, parents' education and interactions of gender and race.  We generate 10,000 samples after the initial 5,000 samples are discarded as a burn-in period, and every 10th sample is saved.  At each MCMC iteration, we computed the Goodman and Kruskal's gamma by generating 2,500 subjects from the posterior predictive distribution.  We observed that the sample paths were stable, and the sample autocorrelations dropped smoothly.  

\begin{figure}[htbp] 
\centering\hspace{-2cm}
\includegraphics[width=17cm, height=21.5cm]{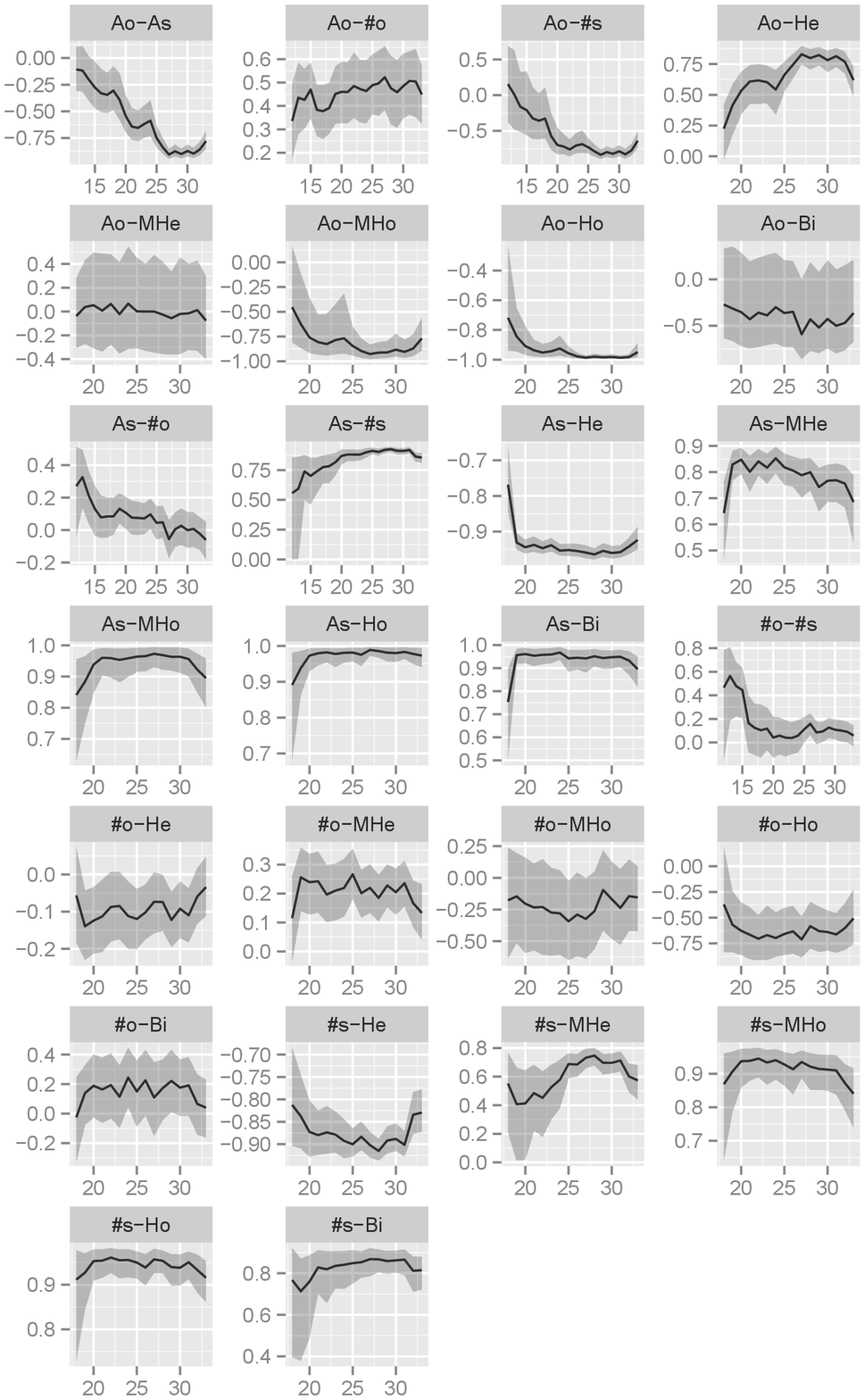}
\caption{Posterior means and 95\% credible intervals of associations for the overall population. $y$-axis shows association and $x$-axis age.}
\label{fig:figure-1}
\end{figure}

We study the association between attraction, behavior, and sexual orientation identity, the latter of which is included starting at age 18, because this variable was not included in wave 1.  Also, we omitted the result for age 11 and 34 from the display below because of the small number of observations, which is 5 and 2 respectively. Figure \ref{fig:figure-1} shows the posterior means and 95\% credible intervals for the gamma statistic in the overall population for opposite sex attraction (Ao), same sex attraction (As), opposite sex partner count (\#o), same sex partner count (\#s), and sexual orientation identities of heterosexual (He), mostly heterosexual (MHe), mostly homosexual (MHo), homosexual (Ho), and bisexual (Bi).  We observe various patterns of trajectories of the associations.  Note that while there is no strong association between the attraction variables in early adolescence, with age a strongly negative association emerges and continues into adulthood.  As one might expect, same sex attraction and same sex partner count have a strong positive association around 0.9, and while the association between opposite sex attraction and partner count is positive, it is more modest, with estimated gamma values around 0.5.  While the association is essentially null in the early teens, with age a strong negative association between opposite sex attraction and same sex partner count becomes apparent. However, the associations between the sexual orientation identity variables (not measured in the early to mid teens) and partner counts tend to be relatively constant over time.

\begin{figure}[htbp] 
\centering\hspace{-2cm}
\includegraphics[width=17cm, height=18cm]{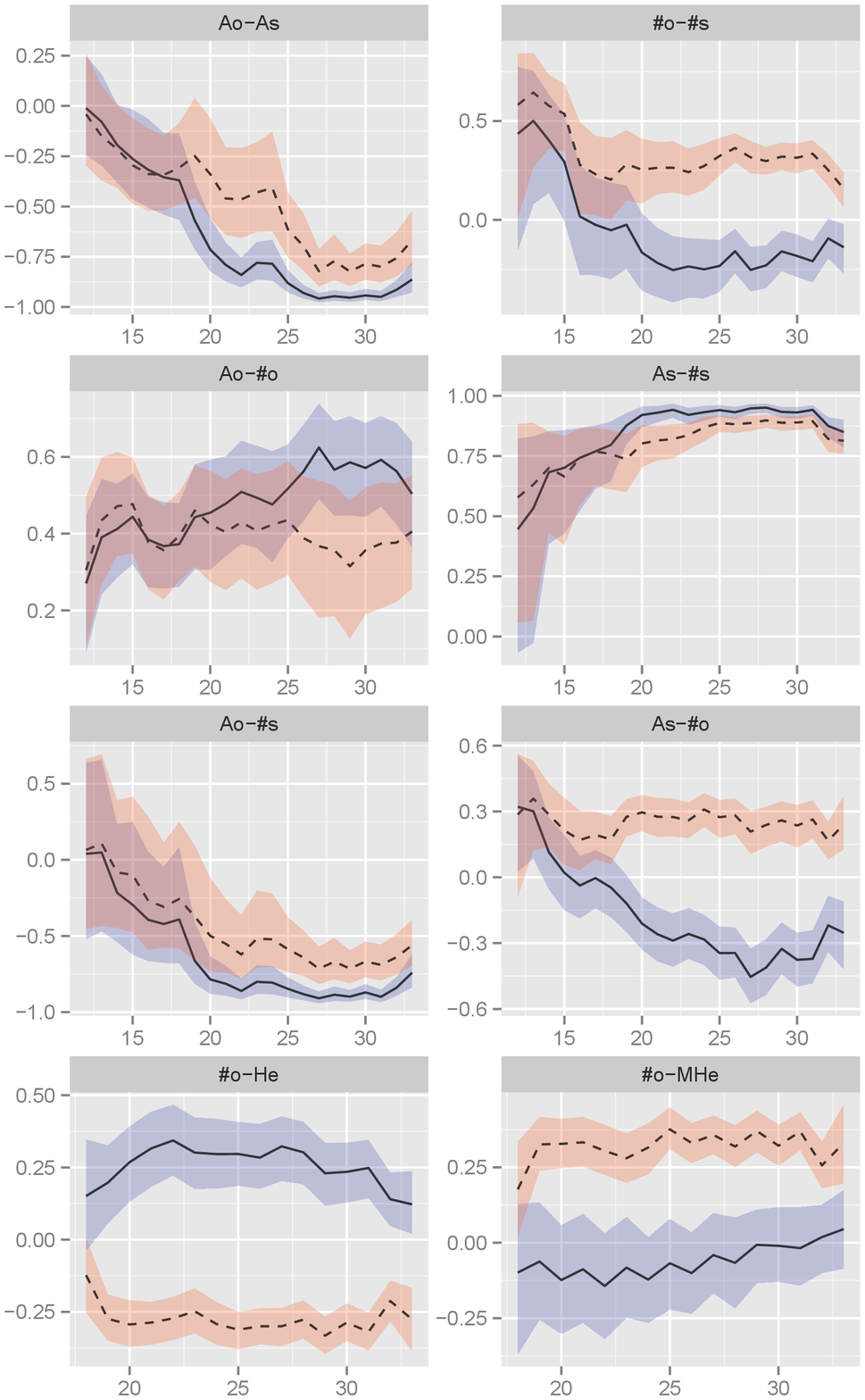}
\caption{Comparison of posterior means and 95\% credible intervals of associations for male (solid line with blue color) and female (dashed line with red color) [1]. $y$-axis shows association and $x$-axis age.}
\label{fig:figure-2}
\end{figure}

\begin{figure}[htbp] 
\centering\hspace{-2cm}
\includegraphics[width=17cm, height=18cm]{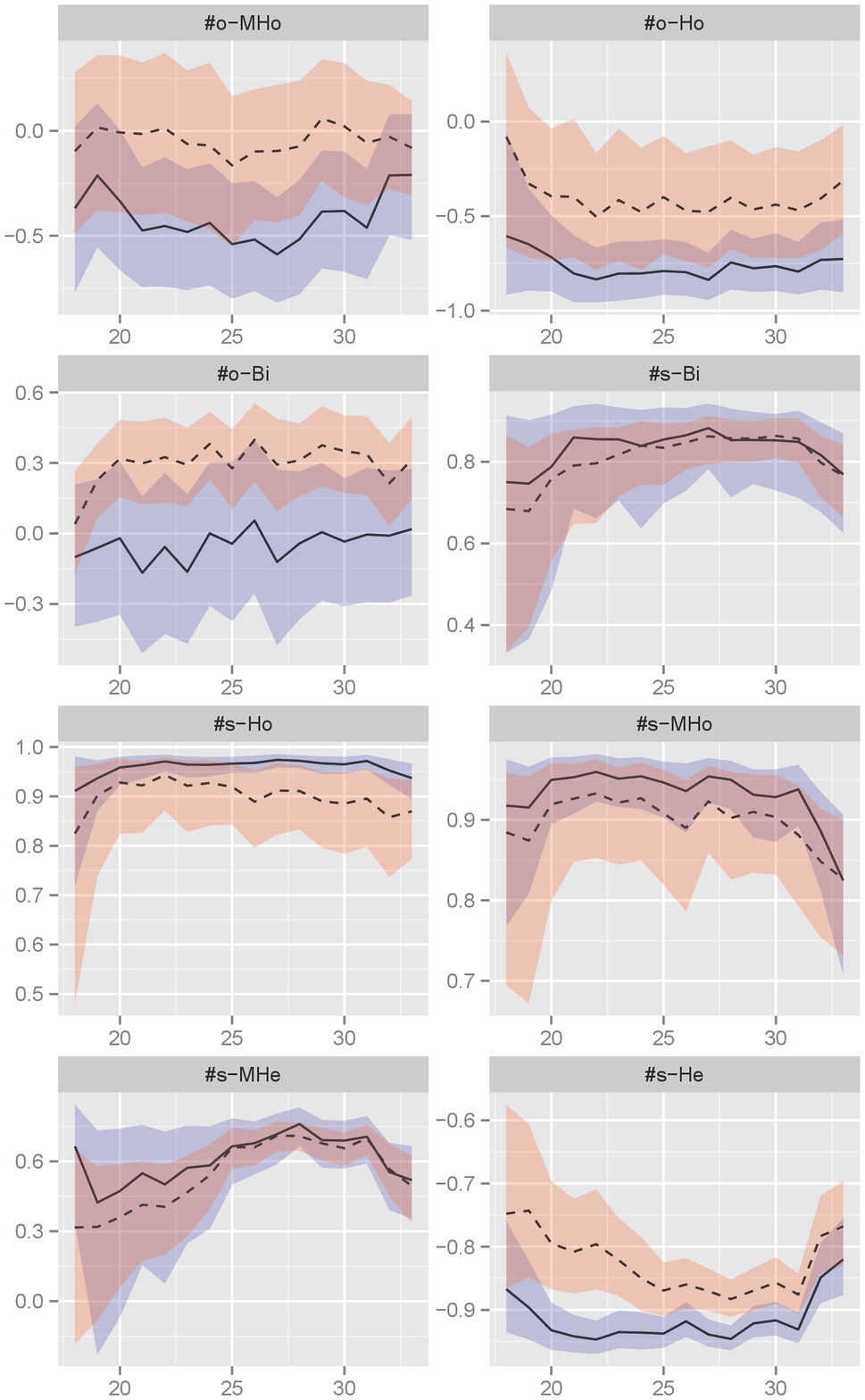}
\caption{Comparison of posterior means and 95\% credible intervals of associations for male (solid line with blue color) and female (dashed line with red color) [2]. $y$-axis shows association and $x$-axis age.}
\label{fig:figure-3}
\end{figure}

While we saw few differences in the associations, or their changes over time, by race, we observed several strong differences by gender. Figures \ref{fig:figure-2} and \ref{fig:figure-3} report the comparison of the posterior means and 95\% credible intervals for several pairs of sexual development variables among males and females.  For example, while the association between counts of same and opposite sex partners is null or positive in the early teens, this association remains significantly positive among females into adulthood but becomes negative among males by the late teens.  We observe a similar trend in the gender-specific associations between same sex attraction and opposite sex partner counts. This pattern is consistent with other findings that lesbian and bisexual-identified females are more likely than heterosexuals to experience both early vaginal intercourse and to engage in sexual risk behaviors such as more sexual partners and inconsistent contraception (\cite{Garofalo98}; \cite{Case04}; \cite{Charlton11}; \cite{McCabe11}; \cite{Herrick13}; \cite{Riskind14}; \cite{McCauley14}; \cite{McCauley15}).  We see opposite trends by gender between heterosexual sexual orientation identity and opposite sex partner counts, with this identity being associated with greater counts among men and lower counts among women from the late teens through early adulthood.  In contrast, a mostly heterosexual identity has a null association with opposite sex partner counts for men, but a positive association among women.  Among both men and women, homosexual identity is associated with lower opposite sex partner counts.  Same sex partner counts are strongly associated with all identities except heterosexual among both men and women; the negative association between heterosexual identity and same sex partner count is stronger among men than among women.

Based on the developmental literature, we also examined whether changes in associations over time varied by gender within race but did not find strong evidence of this interaction in the association parameters.  Similarly, we did not identify strong trends according to parental education. These results are given in the supplementary materials. 
  

\section{Discussion} 

Using data from a longitudinal, naturally-representative sample of adolescents followed into adulthood, we were able to study how the associations between three components of sexual orientation:  attraction, behavior, and sexual orientation identity, evolve with age.  In order to study changes with time in the associations among these variables, measured longitudinally on a variety of categorical scales, we developed a flexible dynamic latent factor model allowing time-varying associations among the multivariate responses.  This model also allowed the associations to depend on covariates of interest, such as gender.  Population-representative inferences are obtained via an MCMC algorithm used for posterior computation.  Previous related work either did not allow longitudinal data (\cite{HoffNiu12}; \cite{FoxDunson15}), required variables to be ordinal (\cite{HoffNiu12}), or did not allow dependence of the association on covariates (\cite{FoxDunson15}).

\bibliographystyle{chicago}
\bibliography{ah}

\end{document}